\documentclass[fleqn,usenatbib]{mnras}

\usepackage{newtxtext,newtxmath}
\usepackage[T1]{fontenc}

\DeclareRobustCommand{\VAN}[3]{#2}
\let\VANthebibliography\thebibliography
\def\thebibliography{\DeclareRobustCommand{\VAN}[3]{##3}\VANthebibliography}

\usepackage{graphicx}	% Including figure files
\usepackage{amsmath}	% Advanced maths commands
\newcommand{\epjc}{Eur. Phys. J. C}

\title[Horndeski gravity with S2]{Testing Horndeski gravity with S2 star orbit}

\author[R. Della Monica et al.]{R. Della Monica,$^{1}$\thanks{E-mail: rdellamonica@usal.es} I. de Martino,$^{1}$ D. Vernieri$^{2,3,4}$ and M. de Laurentis$^{2,3}$
\\
$^{1}$Universidad de Salamanca, Departamento de Fisica Fundamental, P. de la Merced S/N, Salamanca, ES\\
$^{2}$Dipartimento di Fisica, Universit\'a di Napoli {}``Federico II'', Compl. Univ. di Monte S. Angelo, Edificio G, Via Cinthia, I-80126, Napoli, Italy\\
$^{3}$INFN Sezione  di Napoli, Compl. Univ. di Monte S. Angelo, Edificio G, Via Cinthia, I-80126, Napoli, Italy\\
$^{4}$Scuola Superiore Meridionale, Largo San Marcellino 10, I-80138, Naples, Italy
}

\date{Accepted XXX. Received YYY; in original form ZZZ}

\pubyear{2022}

\begin{document}
\label{firstpage}
\pagerange{\pageref{firstpage}--\pageref{lastpage}}
\maketitle

\begin{abstract}
	We have explored a completely new and alternative way to restrict the parameter space of Horndeski theory of gravity. Using its Newtonian limit, it is possible to test the theory at a regime where, given its complexity and the small magnitude of the expected effects, it is poorly probed. At Newtonian level, it gives rise to a generalized Yukawa-like Newtonian potential which we have tested using S2 star orbit data. Our model adds five parameters to the General Relativity model, and the analysis constrains two of them  with unprecedented precision to these energy scales, while only gives an exclusion region for the remaining parameters. We have shown the potential of weak-field tests to constrain Horndeski gravity opening, as a matter of fact, a new avenue that deserves to be further, and deeply, explored near in the future.
\end{abstract}


%%%%%%%%%%%%%%%%%%%%%%%%%%%%%%%%%%%%%%%%%%%%%%%%%%

%%%%%%%%%%%%%%%%% BODY OF PAPER %%%%%%%%%%%%%%%%%%

\section{Introduction}

In the last decades, several extensions of General Relativity (GR) have been proposed in order to overcome its historical limitations, by adding extra ingredients to the gravitational action. Indeed taking into account higher-order derivatives or extra dynamical fields enriches the dynamics and allows to get new phenomenology.
Hereafter, we will consider the possibility to add one extra scalar degree of freedom directly coupled to the metric tensor. In this context we will take into account the very well-known and general class of Horndeski theory of gravity~\citep{Horndeski_1974,Kobayashi_2011} which has been extensively investigated so far.
Such theories have been proposed as the most general ones leading to second order field equations for the metric tensor and the scalar field, even if they have also been extended to more general schemes without increasing the number of degrees of freedom~\citep{Horndeski_1974}.
Horndeski theory has been studied both in cosmological and astrophysical frameworks (for a comprehensive review see \citet{Kobayashi_2019} and the references therein). In the latter case several studies have been conducted in particular concerning black holes~\citep{Maselli_2015,Babichev_2017} and neutron stars~\citep{Cisterna_2016,Maselli_2016}.

The gravitational action of Horndeski theory takes the following form~\citep{Kobayashi_2011}:
\begin{equation}\label{eqn:action}
S = \sum_{i = 2}^{5}\int d^4x \sqrt{-g}\,\mathcal{L}_i[g_{\mu\nu},\phi] + S_m[g_{\mu\nu},\chi_m]\,,
\end{equation}
where $S_m$ is the matter action and $\chi_m$ collectively denotes all matter fields. The gravitational part of the action, which depends on the metric $g_{\mu\nu}$ and one scalar field $\phi$, is given by the following terms:
\begin{align}
\mathcal{L}_2 =& \, G_2(\phi,X)\,, \quad
\mathcal{L}_3 = -G_3(\phi,X)\square\phi\,,\nonumber\\
\mathcal{L}_4 =& \, G_4(\phi,X)R + G_{4X}(\phi,X)\left[(\square\phi)^2 - (\nabla_{\mu}\nabla_{\nu}\phi)^2\right]\,,\nonumber\\
\mathcal{L}_5 =& \, G_5(\phi,X)G_{\mu\nu}\nabla^{\mu}\nabla^{\nu}\phi -\frac{1}{6}G_{5X}(\phi,X)\biggl[(\square\phi)^3 \nonumber\\ 
& -3(\square\phi)(\nabla_{\mu}\nabla_{\nu}\phi)^2 + 2(\nabla_{\mu}\nabla_{\nu}\phi)^3\biggr]\,.\label{eqn:lagrangian_terms}
\end{align}

Generally speaking, the functions \(G_2, \, G_3, \, G_4\), and \(G_5\) are free functions of the scalar field \(\phi\) and its kinetic term \(X\). Each choice of these functions determines a distinct gravity theory which is determined once a suitable mapping is imposed.

Notice that it is possible to map within the general class of Horndeski theories not just those with a scalar field appearing explicitly in the gravitational action and in the corresponding field equations, but also theories (like $f(R)$ gravity or $f(\cal{G})$ Gauss--Bonnet gravity) that inherits higher-order field equations but that somehow admit an equivalent scalar field representation~\citep{Kobayashi_2011}. Additionally, the full Newtonian and Post-Newtonian expansions of the theory have been performed in 2015~\citep{Homann_2015}, leading to an effective general Yukawa-like potential.
Nevertheless, no test of the weak-field limit of Horndeski theory has been made so far by using such a potential, which can indeed allow to constrain the theory at astrophysical scale where it is not yet probed. Hence, we will focus on using the orbital data of the S2 star \citep{Schodel2002, Ghez2003, Eisenhauer2003} orbiting the supermassive black hole (SMBH) in the Galactic Center \citep{Ghez1998, Genzel2010} to bound potential deviation from GR into the Horndeski gravity framework. Although the approach is not new~\citep{Hees2017, deMartino_2021,DellaMonica_2021, DellaMonicaBB}, it has never been applied to such a large class of theories and, therefore, it represents a path never explored that can potentially lead to new bounds on the underlying gravitational theory. More in general, the Yukawa interaction arises in many different scenarios from unification theories \citep{Kaplan2000} to braneworld scenarios \citep{Dvali2000}, modified gravity \citep{deMartino_2021,DellaMonica_2021} and certain models of dark matter \citep{Carroll2009,Carroll2010}. \citet{Hees2017} used the phenomenological framework of the fifth force to constrain the coupling strength and the range or mass of the scalar interaction. Indeed,  using S-stars orbiting around the SMBH in the Galactic Center, they bounded the absolute value of the strength of the fifth force to be $< 0.016$ on a scale length of $\sim 150$ AU. Although mapping these constraints in a specific model is not always a straightforward task, \citet{deMartino_2021} found them to agree with constraints of the Yukawa interaction in $f(R)$-gravity.

This article is organized as follows: in Sec. \ref{sec:horndeski} we introduce the generalized modified Newtonian limit of Horndeski gravity; in Sec. \ref{sec:orbitalmodel} we describe the orbital model for the S2 star in this potential and provide some toy models to illustrate the effects arising in this framework; in Sec. \ref{sec:data} we give a description of the set of observational data used in order to constrain the additional parameters of the theory and our data analysis procedure; in Sec. \ref{sec:results} we report the results of our analysis and finally in Sec. \ref{sec:conclusions} we give our final remarks.

\section{Generalized modified Newtonian potential}
\label{sec:horndeski}

The Newtonian limit of Horndeski gravity theory is calculated through 
a perturbative expansion of the field equations around a Minkowski background
\(\eta_{\mu\nu}\) and a constant cosmological background value \(\Phi\) of the scalar field~\citep{Homann_2015}:
\begin{equation}\label{eqn:metric}
g_{\mu\nu} = \eta_{\mu\nu} + h_{\mu\nu}\,, \quad \phi = \Phi + \psi\,, \quad X = -\frac{1}{2}\nabla^{\mu}\psi\nabla_{\mu}\psi\,.
\end{equation}
Besides assuming that the background is homogeneous and isotropic, we thus also assume that it is stationary,
and additionally that the
four free functions \(G_2, \, G_3, \, G_4\), and \(G_5\), can be expanded in Taylor series as
\begin{equation}\label{eqn:taylorseries}
G_i(\phi,X) = \sum_{m,n = 0}^{\infty}G_{i(m,n)}\psi^mX^n\,,
\end{equation}
where the coefficients \(G_{i(m,n)}\) are given by
\begin{equation}\label{eqn:taylorcoeff}
G_{i(m,n)} = \frac{1}{m!n!}\left.\frac{\partial^{m + n}}{\partial\phi^m\partial X^n}G_i(\phi,X)\right|_{\phi = \Phi, X = 0}\,,
\end{equation}
where each term \(G_{i(m,n)}\psi^mX^n\) is of the order \(\mathcal{O}(\psi^{m+2n})\). 
Notice that, in light of the multi-messenger event GW170817/GRB170817 A, the surviving theories are those for which
\(G_{4,X}=0 \textrm{ and } G_5=0\)~\citep{Creminelli2017}. Additionally, it is worth to notice that  $G_{4,X}$ and $G_5$ do not appear at Newtonian order (we refer the reader to Eqs. (28) in \citet{Homann_2015}).

For a mass point source the solution at Newtonian order is~\citep{Homann_2015}
\begin{equation}\label{eqn:h00def}
h^{(2)}_{00}(r) = \frac{M}{4\pi r}\left[c_2 + \frac{c_1c_{\psi}}{m_{\psi}^2}(e^{-m_{\psi}r} - 1)\right]\,,
\end{equation}
where the constants have been defined as follows:
\begin{align}
c_1 &= -2\frac{G_{4(1,0)}G_{2(2,0)}}{G_{4(0,0)}\Upsilon}\,, \quad
c_2 = \frac{1}{2G_{4(0,0)}} + \frac{G_{4(1,0)}^2}{2G_{4(0,0)}^2\Upsilon}\,,\label{eqn:c2def}
\end{align}
\begin{equation}\label{eqn:mpsi}
m_{\psi} = \sqrt{\frac{-2G_{2(2,0)}}{\Upsilon}}\,, \quad c_{\psi} = \frac{G_{4(1,0)}}{2G_{4(0,0)}}\Upsilon^{-1}\,,
\end{equation}
where for sake of simplicity we have introduced the following definition
\begin{equation}
\Upsilon \equiv G_{2(0,1)} - 2G_{3(1,0)} + 3\frac{G_{4(1,0)}^2}{G_{4(0,0)}}\,.
\end{equation}
Following the seminal work of \citet{Kobayashi_2011}, and the more recent results of \citet{Homann_2015}, Eq.~\eqref{eqn:mpsi} requires $\Upsilon>0$ which implies $G_{2(0,0)}<0$ reducing the allowed parameter space of the theory.
Replacing Eqs.~\eqref{eqn:c2def} and~\eqref{eqn:mpsi} in the metric coefficient~\eqref{eqn:h00def}, one can express the time component of the metric perturbation in the following form:
\begin{equation}\label{eqn:h00}
    h_{00}^{(2)}(r) = \frac{M}{4\pi r}\left[\frac{1}{2 G_{4(0,0)}}+\frac{1}{2}\left(\frac{G_{4(1,0)}}{G_{4(0,0)}}\right)^2\Upsilon^{-1}e^{-m_{\psi}r}\right]\,,
\end{equation}
where $G_{4(0,0)}$ plays the role of the inverse  of gravitation constant $G$.
In a general fashion, Horndeski theory of gravity has a unique limit that recovers GR. 
In particular, by looking at the gravitational action in Eqs.~\eqref{eqn:action} and~\eqref{eqn:lagrangian_terms} it is clear that GR ($\mathcal{L}=R/16\pi G$) is recovered if and only if all the functions $G_2$, $G_3$, and $G_5$ are identically zero (and hence their Taylor expansion coefficient are zero, as well), and $G_4 = (16\pi G)^{-1}$. This is the only physical limit to GR as it involves the cancellation of the additional scalar field $\phi$ introduced to describe the gravitational interaction. On the other hand, at a metric level, the Newtonian approximation of GR for the metric of space-time around a massive point source is given by:
\begin{equation}
    h_{\mu\nu}^{(2)}(r)= \textrm{diag}\left(\frac{2GM}{r}, 0, 0, 0\right).
    \label{eqn:h00GR}
\end{equation}
If we compare Eq.~\eqref{eqn:h00} with Eq.~\eqref{eqn:h00GR}, we see that we can recover the expression of GR by putting $G_{4(0,0)} = 1/16\pi G$ and performing one of the following limits:
\begin{description}
    \item[(A)] $G_{4(1,0)}\to 0$ which recovers Eq.~\eqref{eqn:h00GR} for any value of the other Taylor coefficients;
    \item[(B)] $G_{2(2,0)}\to\pm\infty$ which brings $m_\psi\to \infty$ and hence the cancellation of the Yukawa-like term in Eq.~\eqref{eqn:h00};
    \item[(C)] Any combination of the coefficients $G_{2(0,1)}$, $G_{3(1,0)}$ and $G_{4(1,0)}$ which brings $\Upsilon\to 0$ and, hence $m_\psi \to \infty$, driving the entire Yukawa-like term to an indeterminate form of the kind $e^{-\infty}/0$ whose limit is 0;
    \item[(D)] The simultaneous limit $G_{2(0,1)}$, $G_{2(2,0)}$, $G_{4(1,0)}$, $G_{3(1,0)}, \to 0$ which, again, recovers the expression in Eq.~\eqref{eqn:h00GR} for the metric coefficient.
\end{description}
It is evident that the presence of multiple GR limits, among which only the case (D) corresponds to the physical limit of GR of the theory itself, is a spurious effect due to having performed the Newtonian approximation. Care must be taken, hence, in fixing the values of the coefficients in Eq.~\eqref{eqn:h00} as only one region of the parameter space corresponds to being close to GR, even though at a metric (and thus geodesic) level multiple regions result in a GR-like orbit. The above considerations involve notable degenerations in the parameter space, as different combinations of these parameters return the same physical limit, which cannot be broken without being able to impose {\em a priori} limits on these parameters.

\section{Orbital model}
\label{sec:orbitalmodel}

Let us start summarizing here the technique used to build the orbit of S2 in Horndeski theory of gravity. From the line element in Eqs.~\eqref{eqn:metric} and~\eqref{eqn:h00}, one can compute the geodesic equations, Eqs. (\ref{eqn:ddott}-\ref{eqn:ddotphi}), that describe the motion of test particles in the Newtonian limit.

These equations can be integrated once initial conditions are assigned for the four space-time coordinates and the components of the initial tangent vector to the geodesic. In doing so, one has to take into account the normalization condition for the four-velocity.
Since the metric coefficients do not explicitly depend on the time coordinate $t$, the particular value of $t(0)$ does not affect the integrated orbit. For the S2 star we set for convenience $t(0) = t_A$, where $t_A$ corresponds to the time of the last apocentre passage of the star, which occurred in $\sim 2010$.
Finally, as regards the initial condition on $r$, $\phi$ and the corresponding components of the velocity, we set values that are consistent with the choice $t(0)$. Namely, we set the radial and angular position of the star and its velocity at apocentre. These quantities are estimated by computing the Keplerian osculating ellipse at apocentre, as given by the Keplerian elements measured for S2 in \citet{Gravity_2020} and reported in Table \ref{tab:priors}. 

Once geodesics have been integrated numerically,
we have projected both spatial and velocity vectors of the mock orbit 
in the reference frame of a distant observer in order to be able to compare it with the observational data.
Additionally, in order to properly reconstruct synthetic orbits for S2, we have to take into account on the projected quantities the classical Rømer delay and the redshift on the radial velocity coming from Special Relativity and gravitational time dilation (we refer to \citet{Grould2017, DellaMonica_2021, deMartino_2021, DellaMonicaBB} for a detailed analysis on how these effects can be quantified on our predicted orbits). Finally, we correct our synthetic orbits for systematic effects arising from the construction of the reference frame. More details on the integration procedure we followed can be found in Appendix \ref{sec:numerical},
Thus, our baseline model has 18 parameters, namely: six orbital parameters (the time of pericenter passage $t_p$, the semi-major axis of the orbit $a$, its eccentricity $e$, the inclination $i$, the angle of the line of nodes $\Omega$ and argument of the pericenter $\omega$), two source parameters that are the mass of the central object $M_{\bullet}$ and its distance  $R_{\bullet}$), five system reference parameters, {\em i.e.} ($x_{\bullet}$,$y_{\bullet}$) related to a zero-point offset of the central mass w.r.t. the origin of the astrometric reference frame; ($v_{x,\bullet}$,$v_{y,\bullet}$) describing a drift over time of the central mass on the astrometric reference frame; $v_{z,\bullet}$ related to an offset in the estimation of the radial velocity of the S2 star when correcting to the Local Standard of Rest. Finally, we have five  extra parameters from the expansion of Horndeski gravity at Newtonian order: $G_{2(2,0)}$, $G_{2(0,1)}$, $G_{3(1,0)}$, $G_{4(0,0)}$, and, $G_{4(1,0)}$.

\subsection{Toy models}

\begin{figure*}
    \centering
    \includegraphics[width = \textwidth]{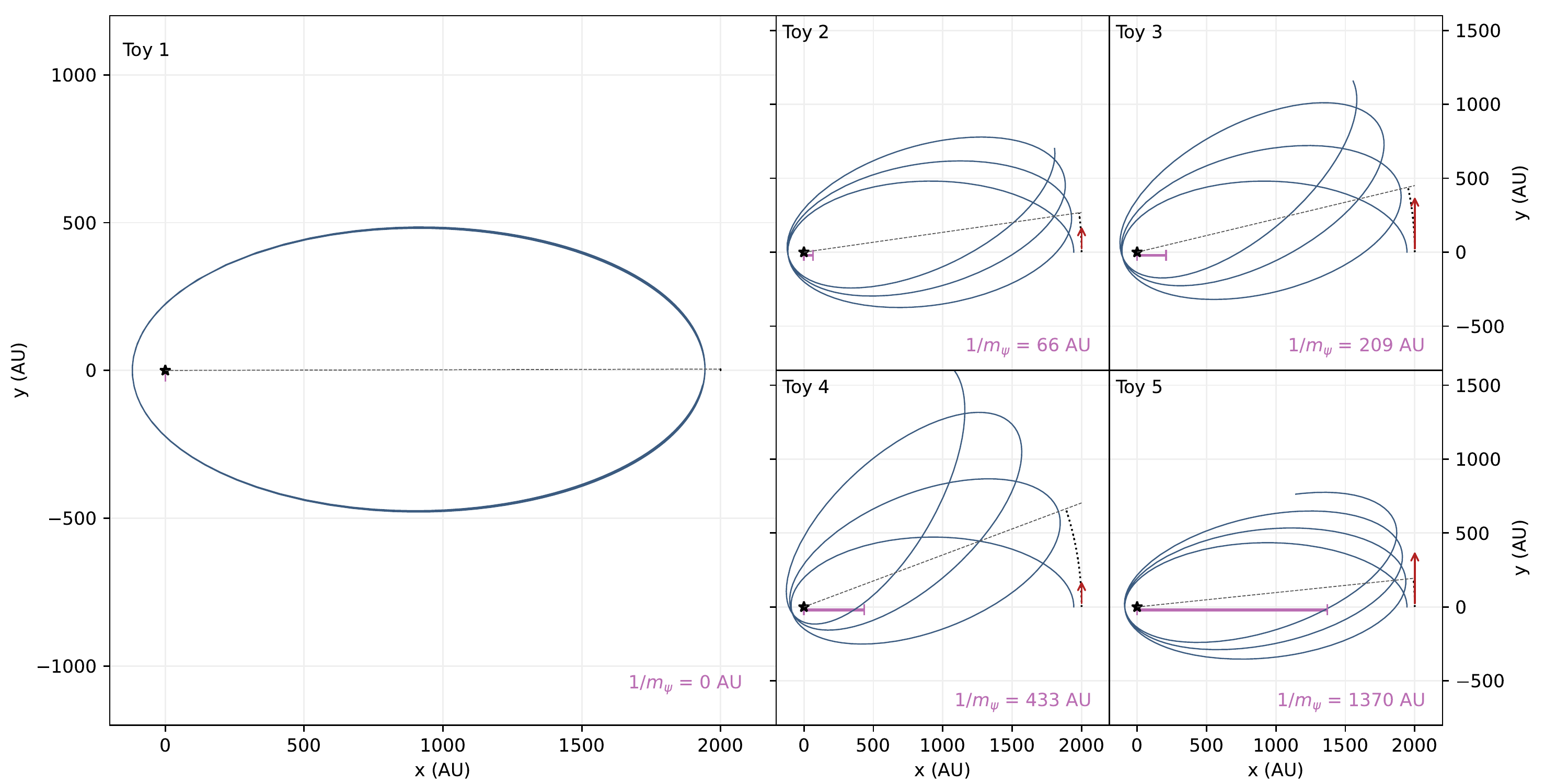}
    \caption{The integrated orbits for the toy models in Table \ref{tab:toy_models}. The pink segment illustrates the varying scale length $1/m_\psi$. The plots report how modification of the orbit are predicted as the coefficient of the  Newtonian expansion of Horndeski are varied. The most noticeable effect that we can visually see on the plots is the pericentre shift on the orbital plane of the star. The quantity $1/m_\psi$ behaves like a scale and it gets more prominent for smaller values of $1/m_\psi$.}
    \label{fig:toy_models}
\end{figure*}

As a first step to study the motion of test particles in the Newtonian expansion of Horndeski gravity for a spherically symmetric space-time, we have selected five toy models for S2 with the same orbital parameters (the ones in Table \ref{tab:priors}) and with varying values for the Horndeski Taylor coefficients. The choice of such parameters has been made in order to assess the deformation of the orbit as a function of $1/m_\psi$ (which has dimensions of a length) and $G_{4(1,0)}$ which, together, modulate the strength and scale length of the modification to the Newtonian potential in Eq. \eqref{eqn:metric}. In particular, we have fixed the GR value of $G_{4(0,0)} = c^4/16\pi G_N$, while we have chosen values for the parameters $G_{4(1,0)}$, $G_{3(1,0)}$, $G_{2(0,1)}$, $G_{2(2,0)}$ that would give a value of $1/m_\psi$ that is a given fraction of the typical scale length of the S2 orbit (e.g. its semi-major axis, $a \sim 1000$ AU). The chosen values are reported in Table \ref{tab:toy_models} and the corresponding orbits from the numerical integration are shown in Figure \ref{fig:toy_models}. In particular, the model Toy 1 represent a star moving under the influence of an unperturbed Newtonian potential, due to the fact that all parameters have been set to zero. Models Toy 2 and Toy 4 have $G_{2(2,0)} = -0.01\;M_\odot /\textrm{AU}\textrm{s}^2$ while for Toy 3 and Toy 5 a value of $G_{2(2,0)} = -0.001\;M_\odot /\textrm{AU}\textrm{s}^2$ has been chosen. On the other hand, while Toy 2 and Toy 3 have a value $G_{4(1,0)} = 10 \;M_\odot \textrm{AU}/\textrm{s}^2$, a value of $G_{4(1,0)} = 100\;M_\odot \textrm{AU}/\textrm{s}^2$ has been fixed for Toy 4 and Toy 5. This choice results in an increasingly high scale length $1/m_\psi$ for the different models, associated with a strength of the Yukawa-like potential in Eq. \eqref{eqn:metric}, modulated by $G_{4(1,0)}$, that is increasingly higher. As can be seen from the integrated orbits in Figure \ref{fig:toy_models}, the main effect related to the variation of the length $1/m_\psi$ is a modification of the orbit and, in particular, its pericentre shift on the orbital plane.
\begin{table*}
    \setlength{\tabcolsep}{10pt}
    \renewcommand{\arraystretch}{1.5}
    \caption{The set of parameters chosen to build our toy models. The second column reports the value of the scale length of the Horndeski modification to the gravitational potential at Newtonian order. Columns 3 to 7, on the other hand, provide with the particular set of parameters resulting in the chosen value of $1/m_\psi$.}
    \centering
    \begin{tabular}{|l|cccccc|}
        \hline
        Model & $1/m_\psi$ & $G_{4(0,0)}$ & $G_{4(1,0)}$ & $G_{3(1,0)}$ & $G_{2(0,1)}$ & $G_{2(2,0)}$\\
         & (AU) & ($M_\odot \textrm{AU}/\textrm{s}^2$) & ($M_\odot \textrm{AU}/\textrm{s}^2$) & ($M_\odot \textrm{AU}/\textrm{s}^2$) & ($M_\odot \textrm{AU}/\textrm{s}^2$) & ($M_\odot /\textrm{AU}\textrm{s}^2$) \\ 
         (1) & (2) & (3) & (4) & (5) & (6) & (7) \\
         \hline
        Toy 1 & 0.0 & 8.0940 & 0 & 0 & 0 & 0\\
        Toy 2 & 66.0 & 8.0940 & 10 & 0 & 50 & -0.01\\
        Toy 3 & 208.6 & 8.0940 & 10 & 0 & 50 & -0.001\\
        Toy 4 & 433.4 & 8.0940 & 100 & 0 & 50 & -0.01\\
        Toy 5 & 1370.5 & 8.0940 & 100 & 0 & 50 & -0.001\\\hline
    \end{tabular}
    \label{tab:toy_models}
\end{table*}

\section{Data and data analysis}
\label{sec:data}

In order to explore the parameter space of the coefficients arising in the Newtonian approximation of Horndeski gravity for a spherically symmetric space-time, we have employed publicly available kinematic data for the S2 star. In particular, the data presented in Table 5 of \citet{Gillessen_2017} provide with 145 astrometric data-points of the recorded sky-projected position of S2 with respect to the "Galactic Centre (GC) infrared reference system''~\citep{Plewa_2015} and 44 measurements of its velocity along the line of sight, corrected for the Local Standard of Rest (LSR), that have been retrieved from line shift through spectroscopic observations. These data cover a period spanning from $\sim 1992$ to $\sim 2016$. This means that, given that the period of S2 is $\sim 16$ years, a full orbit is covered, but no data is present for the crucial pericentre passage occurred in May 2018. The data that have been recorded with GRAVITY interferometer during this passage are not publicly available, but have been used to place significant constraints on the first post-Newtonian order of GR ~\citep{Gravity_2018, Gravity_2020}. An independent analysis by the Galactic Center group at the University of California, Los Angeles, was able to detect the gravitational redshift at the S2, called S-02 in their works, pericenter using data collected at the Keck observatory \citep{Do2019}. Nevertheless, the available data are sufficient for our purpose, since our aim is to perform for the first time an exploration of the parameter space for Horndeski gravity at Newtonian order. 

We have employed the Bayesian sampling of the posterior probability distribution of these parameters treated as random variables. In particular this was done by performing a Markov Chain Monte Carlo (MCMC) analysis, implemented in the \texttt{emcee}~\citep{emcee} module. In order not to force the results to a Keplerian orbit, we have not fixed the orbital elements of S2 to the values derived in previous analysis on the same dataset~\citep{Gillessen_2017}, but we have left them as free parameters of our fit.  Thus, the sampler draws random values for all the 18 aforementioned parameters, within appropriate prior distributions, and, for each parameter-set, we perform a synthetic orbit for S2, as illustrated above. The agreement between the numerically integrated orbit and the observational data is assessed with the following log-likelihood function:
\begin{align}
	-2\log\mathcal{L} =& \sum_i\biggl[\biggl(\frac{X_{\rm obs}^i-X^i}{{\sigma_{X,{\rm obs}}^i}}\biggr)^2+
	\biggl(\frac{Y_{\rm obs}^i-Y^i}{{\sigma_{Y,{\rm obs}}^i}}\biggr)^2+\biggl(\frac{V_{\rm Z, obs}^i-V_Z^i}{{\sigma_{V_Z,{\rm obs}}^i}}\biggr)^2\biggr]\,,
\end{align}
where ($X_{\rm obs}$, $Y_{\rm obs}$, $V_{Z, \rm obs}$) are the observed positions and radial velocities for S2, while ($X$, $Y$, $V_Z$) are the ones from our numerical integration. Convergence of the algorithm is assessed with the estimation of the autocorrelation time of the Markov chains~\citep{Goodman_2010}. The particular priors used for our analysis are given by a set of univariated Gaussian distributions on the orbital parameters whose central values are fixed on the best-fitting values of \citet{Gillessen_2017}  and whose full weight half maximum are chosen to be ten times the uncertainty on these parameters from \citet{Gillessen_2017} (reported in Table \ref{tab:priors}). The parameters related to the systematic effects in the construction of the reference frame have been assigned Gaussian priors from an independent analysis in \citet{Plewa_2015}. Finally, for the five parameters of the Newtonian expansion of Horndeski gravity we used uniform priors whose amplitudes have been set heuristically. All priors are listed in Table \ref{tab:priors}.

\begin{table*}
	\setlength{\tabcolsep}{13.5pt}
	\renewcommand{\arraystretch}{1.7}
	\caption{The full set of orbital parameters for the orbit of the S2 star as estimated by \citet{Gillessen_2017} used as priors of our analysis (for there parameters we considered gaussian distributions whose FWHM are 10 times the experimental errors reported here) and the reference frame priors on $x_\bullet$, $y_\bullet$, $v_{x,\bullet}$, $v_{t,\bullet}$ and $v_{z,\bullet}$ from \citet{Plewa_2015}. In the bottom-right part of the table we report the set of priors used for the parameters $G_{4(0,0)}$, $G_{4(1,0)}$, $G_{3(1,0)}$, $G_{2(0,1)}$ and $G_{2(2,0)}$ in our analysis. We have left $G_{4(0,0)}$ vary around its GR value in an interval of amplitude $10\%$. The other intervals have been set heuristically, starting from some guessed values and than widening or narrowing the interval as needed. The parameter $G_{2(2,0)}$ has been fixed to be negative and thus we assumed an additional prior to set $\Upsilon > 0$ \citep{Kobayashi_2011, Homann_2015}.}
	\begin{tabular}{|lccr||lccr|}
	\hline
	{\bf Parameter} & {\bf Unit} & {\bf Value}   & {\bf Error} & {\bf Parameter} & {\bf Unit} & {\bf Value}   & {\bf Error}  \\ \hline
	$M_\bullet$   & $10^6M_\odot$ & 4.35       & 0.012 & $y_\bullet$ & mas & $0.1$&$0.2$\\
	$R_\bullet$   & kpc           & 8.33      & 0.0093 & $v_{x,\bullet}$ & mas/yr & $0.02$&$0.2$\\
	$a$           & mas           & 125.5     & 0.044 & $v_{y,\bullet}$ & mas/yr & $0.06$&$0.1$\\
	$e$           &                 & 0.8839  & 0.000079 & $v_{z,\bullet}$ & km/s & $0$ & 5\\ \cline{5-8}
	$i$           & $^\circ$      & 134.18     & 0.033 &  &  & \multicolumn{2}{c|}{{\bf Interval}} \\ \cline{5-8}
	$\omega$      & $^\circ$      & 65.51      & 0.030 & $G_{4(0,0)}$                  & $M_\odot \textrm{AU}/\textrm{s}^2$ & \multicolumn{2}{c|}{$[0.95, 1.05](c^4/16\pi G)$} \\
	$\Omega$      & $^\circ$      & 226.94     & 0.031 & $G_{4(1,0)}$                  & $M_\odot \textrm{AU}/\textrm{s}^2$ & \multicolumn{2}{c|}{{[}-100, 100{]}}          \\
	$T$           & yr            & 16.00     & 0.0013 &  $G_{3(1,0)}$                  & $M_\odot \textrm{AU}/\textrm{s}^2$ & \multicolumn{2}{c|}{{[}-100000, 100000{]}}\\
	$t_p$         & yr            & 2018.33  & 0.00017 & $G_{2(0,1)}$                   & $M_\odot \textrm{AU}/\textrm{s}^2$ & \multicolumn{2}{c|}{{[}-100000, 100000{]}}\\
	$x_\bullet$ & mas & $-0.2$&$0.2$ & $G_{2(2,0)}$                   & $M_\odot/ \textrm{AU}\textrm{s}^2$ & \multicolumn{2}{c|}{{[}-10, 0{]}}\\ \hline
	\end{tabular}
	\label{tab:priors}
\end{table*}

\section{Results}
\label{sec:results}

The posterior distributions of the 18 parameters of our orbital model for the S2 star in Horndeski gravity are reported in the online supplementary materials. More specifically, we report the 68, 95, and 99.7\% confidence intervals for all the parameters. It is worth mentioning that all the orbital elements and the reference frame parameters for the S2 star are bounded and recover, within $3\sigma$ (at most), all fiducial values reported in Table \ref{tab:priors}. Only the semi-major axis $a$ and the inclination angle $i$ represent an exception, being correlated with the parameter $G_{4(0,0)}$, and are only marginally compatible with their counterpart obtained in a Newtonian framework in \citet{Gillessen_2017}. This is anyway rather expected since  $G_{4(0,0)}$ modulates the Newtonian gravitational constant $G$ hence affecting the posterior distribution of the proposed values of the semi-major axis and of the inclination angle. 

\begin{figure}
    \centering
    \includegraphics[width = \columnwidth]{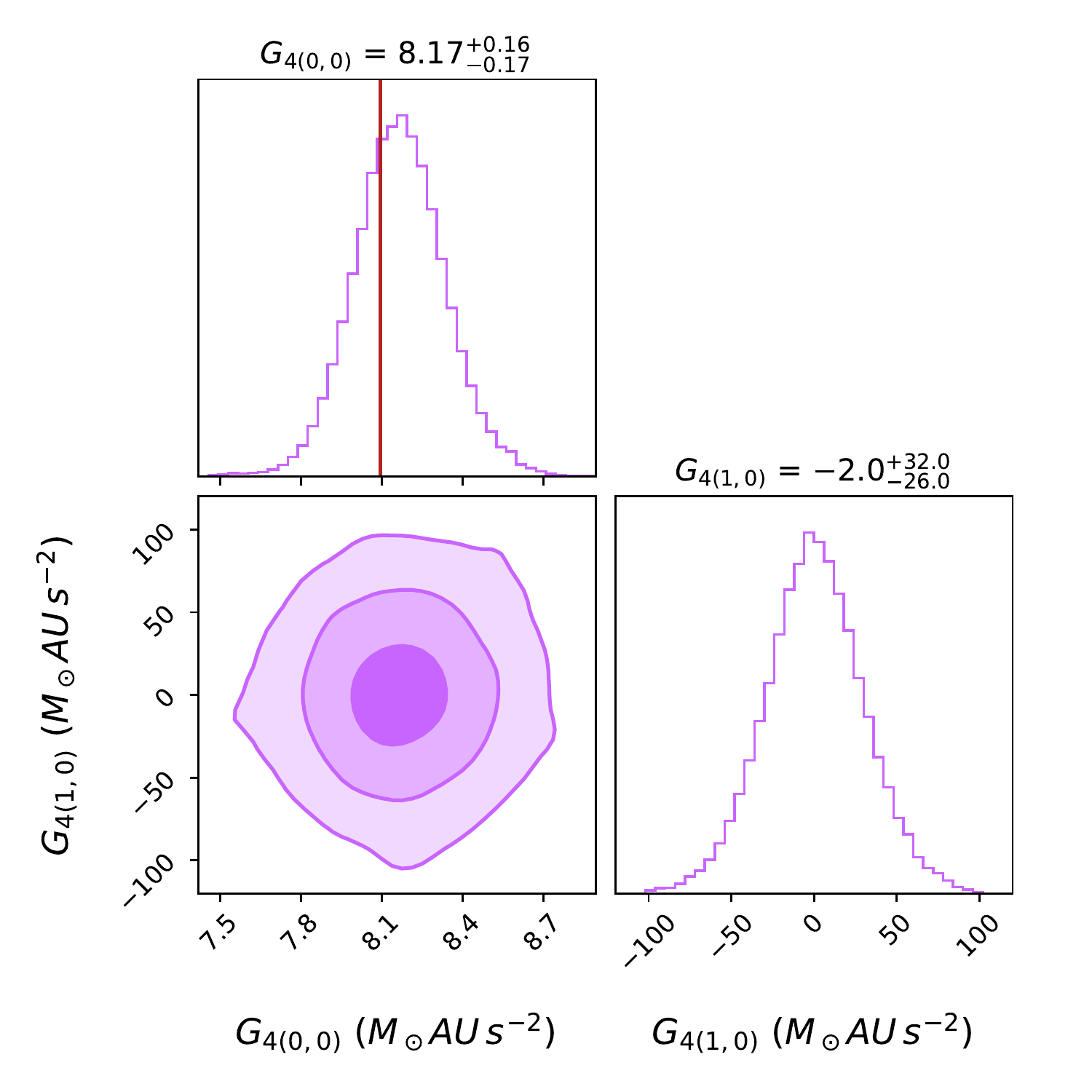}
    \caption{Focus on the posterior distribution of the Taylor coefficients of Horndeski gravity. The vertical red line in the top panel depicts the expected value in GR of the Taylor coefficient $G_{4(0,0)}$. }
    \label{fig:mcmc_posterior_1_crop}
\end{figure}

\begin{figure}
    \centering
    \includegraphics[width = \columnwidth]{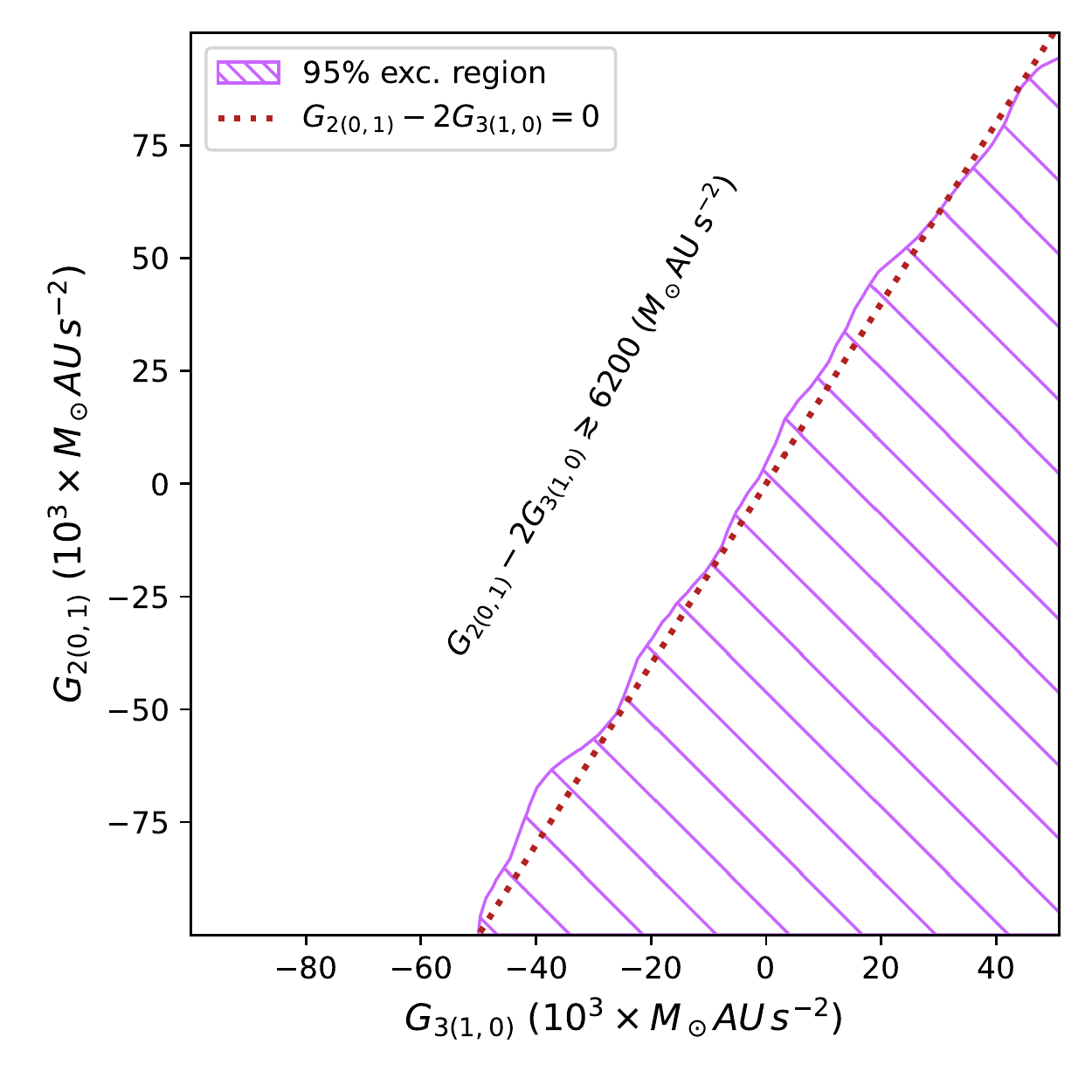}
    \caption{95\% confidence level exclusion regions of the $G_{2(0,1)}$ - $G_{3(1,0)}$ slice of the parameter space from our posterior analysis.}
    \label{fig:exclusion_regions}
\end{figure}

\begin{figure}
    \includegraphics[width = 0.45\textwidth]{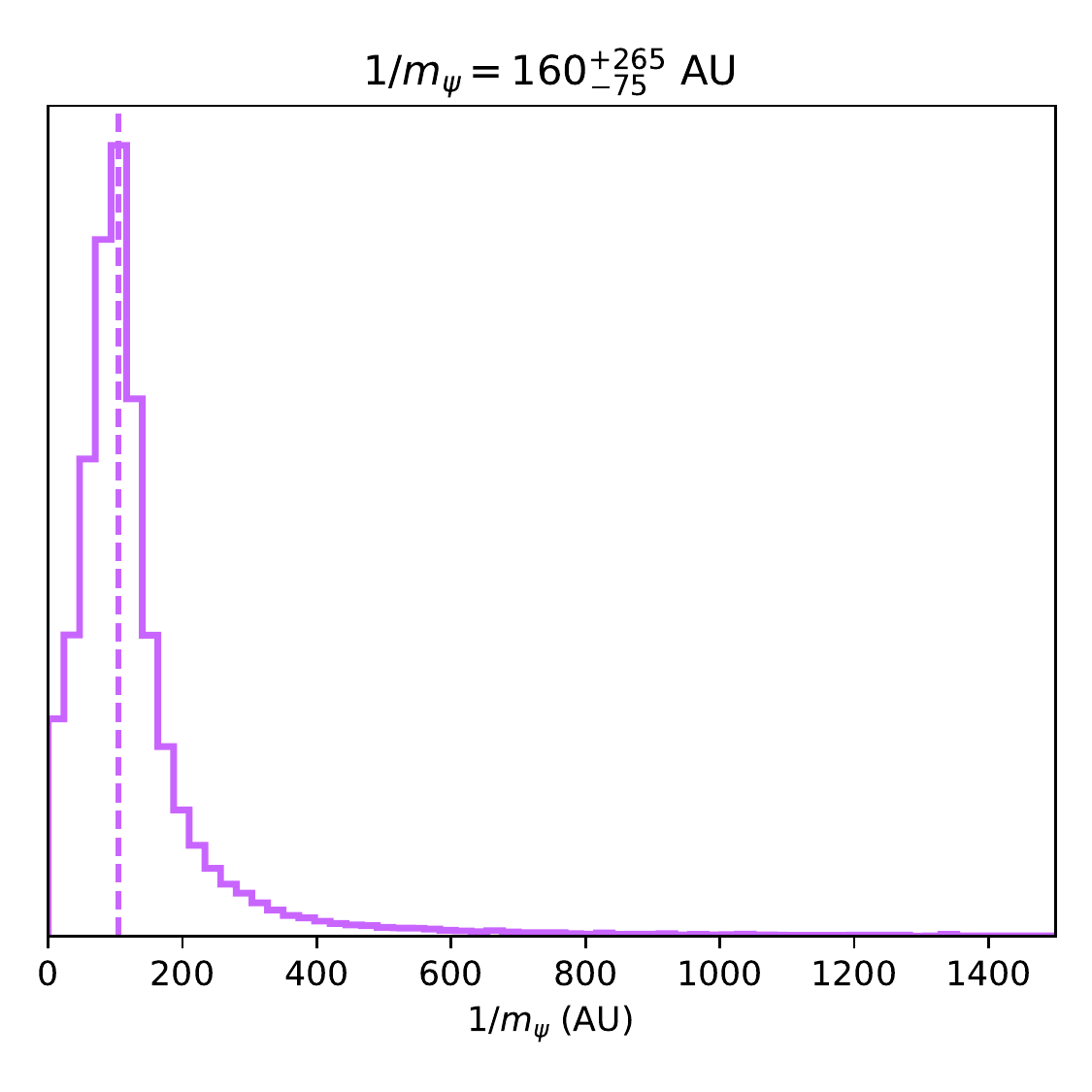}
    \caption{Posterior distribution on the inverse of $m_\psi$ in AU derived from the posteriors of the four parameters in Eq. \eqref{eqn:mpsi}. Dashed vertical lines correspond to the median value of the posterior distribution, also reported in Table \ref{tab:results}.}
    \label{fig:1_mpsi}
\end{figure}

Coming to the additional Horndeski parameters, we place constraint on $G_{4(0,0)}$ and $G_{4(1,0)}$, while the other parameter remains unconstrained. Nevertheless, we were able to place a constraint on a combination of the remaining parameters and determined an exclusion region. In detail, Figure \ref{fig:mcmc_posterior_1_crop} is an inset of the corner plot of the full posterior reporting the 68, 95, and 99.7\% confidence intervals of $G_{4(0,0)}$ and $G_{4(1,0)}$ parameters, and their corresponding posterior distributions. As reported in the Figure, we find $G_{4(0,0)}=8.17^{+0.16}_{-0.17} M_\odot$AU/s$^2$ and $G_{4(1,0)}=-2^{+32}_{-26} M_\odot$AU/s$^2$. The parameter $G_{4(0,0)}$ is fully compatible with its expected value in GR (which is depicted as a vertical red line in the top panel), and the parameter $G_{4(1,0)}$ is compatible with zero at $1\sigma$  reducing the gravitational potential to the Newtonian one: $h_{00}=\frac{M}{8\pi G_{4(0,0)} r}$. The result was rather expected but still surprising because the orbital motion data of S2 star, that we are using to test the Newtonian limit of Horndeski theory of gravity, cannot distinguish GR from Newton theory either~\citep{Gillessen_2017}. Previous tests of the weak field limit of Horndeski gravity strongly constrained $G_{4(0,0)}$, but they consider a subclass of the whole theory by setting $G_{2(2,0)}=0$ which means $m_\psi =0$~\citep{Hou2018}. In this case, the Yukawa-term disappears reducing itself to an additional constant in Eq.~\eqref{eqn:h00}. On the other hand, none was able to place constraints on $G_{4(1,0)}$ in the weak field limit. Additionally to those new bounds, looking at  the contours of the $G_{2(0,1)}$ - $G_{3(1,0)}$ slice of the whole parameter space, we can see that even though we are not able to place a definitive constraint on either one of the two parameters, we are able to exclude a region of the parameter space that results in  a lower limit on the combination $G_{2(0,1)}-2G_{3(1,0)} > 6200 M_\odot$AU/s$^2$ as shown in Figure \ref{fig:exclusion_regions}.  Finally, we were able to bound the scale length $m_\psi^{-1}$. First, we randomly sampled $m_\psi^{-1}$ from the posterior of Horndeski gravity parameters carrying out a 1,000 Monte Carlo simulation to estimate the distribution of $m_\psi$ and its variance. As shown in Figure \ref{fig:1_mpsi}, we found $m_\psi^{-1} = 160 ^{+265}_{-75}$ AU. Noticeably, this represents the best constraint up to date on $m_\psi^{-1}$ since other analyses only place a lower limit $m_\psi^{-1} \geq 9 $ AU using binary systems composed by a pulsar plus a white dwarf~\citep{Dyadina2019}. Finally, let us point out that, even though other constraints exist in literature for a phenomenological Yukawa-like modification to the Newtonian potential using the S2 star at the GC (see e.g. \citet{Hees2017}), due to the specific dependence of Eq. \ref{eqn:h00} on the Taylor parameter of the Horndeski functions, there is no way to consistently map constraints on the former potential into bounds on the latter. Indeed, phenomenological Yukawa-like modifications are parametrized by a scale length ($\lambda$) and a modulation factor ($\alpha$), there are treated (both conceptually and statistically) as independent parameters. The equivalent scale length $1/m_\psi$ in our metric \eqref{eqn:h00}, on the other hand, is not independent from the coefficient ${1}/{2\Upsilon}\left({G_{4(1,0)}}/{G_{4(0,0)}}\right)^2$ of the exponential term in the metric, due to the fact that $m_\psi$ itself depends on all the four free parameters of the Newtonian limit studied here. This, along with the intrinsic non-linearity of the relation between the Horndeski parameters and those in the Yukawa-like potential, makes it conceptually impossible to map pre-existing bounds on $\alpha$ and $\lambda$ into bounds on $G_{4(0,0)}$, $G_{4(1,0)}$, $G_{3(1,0)}$ and $G_{2(0,1)}$.

\begin{table*}
    \setlength{\tabcolsep}{17.5pt}
    \renewcommand{\arraystretch}{1.8}
    \caption{Results of our posterior analysis on all the parameters of our orbital model for the S2 star in the Newtonian expansion of Horndeski's gravity. In particular, we report the 68\% confidence interval for all the orbital and reference frame parameters; the 68\% confidence interval for $G_{4(0,0)}$ (which is bounded and perfectly compatible to the Newtonian value $c^4/16\pi G_N$) and for $G_{4(1,0)}$ (which is centered on 0). The parameter $G_{2(2,0)}$ results unbounded from our analysis, while we are only able to put a constraint on the combination $G_{2(0,1)}-2G_{3(1,0)}$ which appears as an exclusion region on the corresponding contour plot. Finally, we report the 68\% confidence interval for the scale length $1/m_\psi$.}
    \label{tab:results}
    \centering
    \begin{tabular}{|lcr||lcr|}
    \hline
        {\bf Parameter} & {\bf Unit} & {\bf Best-fit} & {\bf Parameter} & {\bf Unit} & {\bf Best-fit} \\ \hline
        $R_\bullet$ & kpc & $8.53^{+0.13}_{-0.12}$ & $y_\bullet$ & mas & $0.2^{+1.8}_{-2.1}$\\
        $T$ & yr & $15.95^{+0.23}_{-0.15}$ & $v_{x,\bullet}$ & mas/yr & $-0.02^{+0.12}_{-0.08}$\\
        $t_p - 2018$ & yr & $0.33^{+0.11}_{-0.08}$ & $v_{y,\bullet}$ & mas/yr & $0.00^{+0.09}_{-0.11}$\\
        $a$ & mas & $0.12323^{+0.00052}_{-0.00048}$ & $v_{z,\bullet}$ & km/s & $7.5^{+3.9}_{-4.0}$\\
        $e$ &  & $0.8821^{+0.0027}_{-0.0027}$ & $G_{4(0,0)}$ & $M_\odot$ AU s$^{-2}$ & $8.17^{+0.16}_{-0.17}$\\
        $i$ & $(\,^\circ)$ & $135.12^{+0.18}_{-0.17}$& $G_{4(1,0)}$ & $M_\odot$ AU s$^{-2}$& $-2.0^{+32.0}_{-26.0}$\\
        $\Omega$ & $(\,^\circ)$ & $226.09^{+0.54}_{-0.54}$ & $G_{2(0,1)}-2G_{3(1,0)}$ &$M_\odot$ AU s$^{-2}$ & $\gtrsim 6200$\\
        $\omega$ & $(\,^\circ)$ & $64.25^{+0.54}_{-0.44}$ & $G_{2(2,0)}$ & $M_\odot$AU$^{-1}$s$^{2}$& Unbounded\\
        $x_\bullet$ & mas & $-0.4^{+1.6}_{-2.4}$ & $1/m_\psi$ & AU & $160^{+265}_{-75}$\\
        \hline
    \end{tabular}
\end{table*}

\section{Discussions and Conclusions}
\label{sec:conclusions}

Horndeski theory of gravity represents the most general scalar-tensor theory (including - but not limited to - Brans--Dicke, $f(R)$ gravity, $f(\cal{G})$ Gauss--Bonnet gravity, covariant Galileons, and chameleons among the others), with an additional scalar degree of freedom coupled to the metric and leading to second order equations of motion. While formulated almost 50 years ago, this model has only recently been back in vogue and extensively investigated at cosmological scales~\citep{Horndeski_1974}. Nevertheless, it lacks a deep investigation in the weak field limit. On one hand, this is due to the fact that a weak field approximation was formulated only a few years ago in 2015~\citep{Homann_2015}. On the other hand, the intrinsic complexity of such a weak field limit, reflecting in a very high number of additional parameters with respect to GR, makes it rather complicated to extract valuable constraints from astrophysical dataset. Indeed, a previous study constraining the weak field limit of Horndeski theory, was only able to bound $G_{4(0,0)}$ in a subclass of models that set $G_{2(2,0)}=0$, which means $m_\psi =0$, using time-delay data from the Cassini mission and gravitational wave radiation from binary pulsar systems~\citep{Hou2018}. The only previous analysis investigating the whole theory uses binary systems to place a lower limit on the parameter $m_\psi^{-1} \geq 9 $ AU~\citep{Dyadina2019}. Therefore, a comprehensive analysis constraining the Taylor coefficients of the functions \(G_2, \, G_3, \, G_4\), and \(G_5\) appearing in the weak field limit in Eq.~\eqref{eqn:h00} was fully missing so far.

Using the orbital motion of the S2 star around the supermassive black hole in the centre of the Milky Way, we have placed constraints on two Taylor coefficients: $G_{4(0,0)}=8.17^{+0.16}_{-0.17} M_\odot$AU/s$^2$ and $G_{4(1,0)}=-2^{+32}_{-26} M_\odot$AU/s$^2$; and on a combination of other two coefficients $G_{2(0,1)}-2G_{3(1,0)} > 6200 M_\odot$AU/s$^2$. These results recover the value of $G_{4(0,0)}$ expected in GR at $1\sigma$, and set a new bound on the parameter $G_{4(1,0)}$ that encodes the deviations from GR. Our constraint shows a full accordance with the GR value of $G_{4(1,0)}=0$ within $1\sigma$. Moreover, we also are capable to set a strong bound on the previously unbounded parameter 
$m_\psi^{-1}$ that tunes the scale on which the novel Yukawa-interaction acts.

This analysis does not only provide with unprecedented constraints on Horndeski gravity at astrophysical scales that had never been tested before; it also hints the potential of future astronomical observations to constrain the theory in both the weak field limit through the Post-Newtonian approximation that can be probed by the forthcoming GRAVITY data~\citep{Gravity2017}, and the strong field limit that can be probed through the black hole solutions in Horndeski gravity using horizon-scale Event Horizon Telescope observations~\citep{Doeleman2008, Broderick2009}. Joint observations exploring both weak and strong field limits will potentially lead to either endorse or rule out higher-order theories of gravity.

\section*{Acknowledgements}

RDM acknowledges support from Consejeria de Educación de la Junta de Castilla y León and from the Fondo Social Europeo.
IDM acknowledges support from Ayuda  IJCI2018-036198-I  funded by  MCIN/AEI/  10.13039/501100011033  and:  FSE  “El FSE  invierte  en  tu  futuro”  o  “Financiado  por  la  Unión  Europea   “NextGenerationEU”/PRTR. 
IDM is also supported by the project PGC2018-096038-B-I00 funded by the Spanish “Ministerio de Ciencia e Innovación” and FEDER “A way of making Europe", and by the project SA096P20 Junta de Castilla y León.
DV and MDL acknowledge the support of Istituto Nazionale di Fisica Nucleare (INFN) {\it iniziative specifiche} QGSKY and TEONGRAV. DV also acknowledges the FCT project with ref. number
PTDC/FIS-AST/0054/2021.

%%%%%%%%%%%%%%%%%%%%%%%%%%%%%%%%%%%%%%%%%%%%%%%%%%
\section*{Data Availability}

Data used in this article are publicly available on the electronic version of \citet{Gillessen_2017} at \url{https://iopscience.iop.org/article/10.3847/1538-4357/aa5c41/meta\#apjaa5c41t5}.


\bibliographystyle{mnras}
\bibliography{biblio} % if your bibtex file is called example.bib

%%%%%%%%%%%%%%%%%%%%%%%%%%%%%%%%%%%%%%%%%%%%%%%%%%

%%%%%%%%%%%%%%%%% APPENDICES %%%%%%%%%%%%%%%%%%%%%

\appendix

\section{Numerical integration of the geodesic equations}
\label{sec:numerical}
The gravitational field around a mass point source is described in Horndeski's gravity at Newtonian order by the line element in Eq.~\eqref{eqn:metric}. From it one can compute the geodesic equations that describe the motion of freely falling test particles in the Newtonian limit and for the four spherical isotropical coordinates:
\begin{align}
    \ddot{t} &= -\frac{M\dot{t}\dot{r}\left[\frac{1}{r}\biggl(G_{4(0,0)}\Upsilon+G_{4(1,0)}^2e^{-m_{\psi}r}\biggr)+G_{4(1,0)}^2m_\psi e^{-m_{\psi}r}\right]}{8\pi G_{4(0,0)}\Upsilon r-G_{4(0,0)}\Upsilon M-G_{4(1,0)}^2M e^{-m_{\psi}r}}\,,\label{eqn:ddott}\\
    \ddot{r}&=\displaystyle r \dot{\phi}^{2} \sin^{2}{\theta} + r \dot{\theta}^{2} + \frac{\dot{t}^{2} M}{16\pi G_{4(0,0)}r}\left[- \frac{1}{r} - \frac{G_{4(1,0)}^{2}  e^{-m_{\psi}r}}{G_{4(0,0)} \Upsilon} \biggl( m_\psi + \frac{1}{ r}\biggr)\right]\,,\label{eqn:ddotr}\\
    \ddot{\theta}&=\displaystyle \dot{\phi}^{2} \sin{\theta} \cos{\theta} - \frac{2 \dot{r} \dot{\theta}}{r} \label{eqn:ddottheta}\,,\\
    \ddot{\phi}&= \displaystyle - \frac{2 \dot{\phi}\dot{\theta} \cos{\theta}}{\sin{\theta}} - \frac{2 \dot{\phi} \dot{r}}{r}\,,\label{eqn:ddotphi}
\end{align}
where a dot represents a derivative with respect to an affine parameter $\tau$.

These equations can be integrated once initial conditions are assigned for the four space-time coordinates and the components of the initial tangent vector to the geodesic. In doing so, one has to take into account the normalization condition for the four-velocity. In particular, for a time-like geodesic, which describes the world-line of a massive particle, it is required that $g_{\mu\nu}\dot{x}^\mu\dot{x}^\nu=-1$. This condition allows to assign the initial value for one of the $\dot{x}^\mu$ components, given the other three.
Furthermore, as it appears from Eq. \eqref{eqn:ddottheta}, if initial conditions are assigned so that $\theta(0) = \pi/2$ and $\dot{\theta} = 0$, a value $\ddot{\theta} = 0$ is obtained, identically. This implies that one can always define the reference frame so that the motion of the test particle occurs in the equatorial plane. 
Moreover, since the metric coefficients do not explicitly depend on the time coordinate $t$, the particular value of $t(0)$ does not affect the integrated orbit. For the S2 star we set for convenience $t(0) = t_A$, where $t_A$ corresponds to the time of the last apocentre passage of the star, which occurred in $\sim 2010$.
Finally, as regards the initial condition on $r$, $\phi$ and the corresponding components of the velocity, we set values that are consistent with the choice $t(0)$. Namely, we set the radial and angular position of the star and its velocity at apocentre. These quantities are estimated by computing the Keplerian osculating ellipse at apocentre, as given by the Keplerian elements \citep{Green1985} measured for S2 from \citep{Gillessen_2017} and reported in Table \ref{tab:priors}. 

We have integrated Eqs. (\ref{eqn:ddott}-\ref{eqn:ddotphi}) numerically, by means of an adaptive step-size Runge-Kutta algorithm \citep{Dormand1980}, obtaining parametric arrays $\{t(\tau), r(\tau), \theta(\tau),\phi(\tau)\}$ and $\{\dot{t}(\tau), \dot{r}(\tau), \dot{\theta}(\tau),\dot{\phi}(\tau)\}$ that describe the motion of S2 in a reference frame centered on the central source of the gravitational field. In order to be able to compare the integrated orbit with the observational data, a projection is required in the reference frame of a distant observer, by means of the following relations:
\begin{align}
     X&=\mathcal{B}x+\mathcal{G}y\,,\\
     Y&=\mathcal{A}x+\mathcal{F}y\,,\\
     Z&=\mathcal{C}x+\mathcal{H}y + D\,.\label{eq:z_obs}
\end{align}
where $(x,y)$ are the coordinates of the star on its orbital plane, while $(X,Y)$ and $Z$ are the sky-projected position of the star and its distance from the observer, respectively. The constants $\mathcal{A}$, $\mathcal{B}$, $\mathcal{C}$, $\mathcal{F}$, $\mathcal{G}$ and $\mathcal{H}$ are called Thiele-Innes elements \citep{Taff1986} and are obtained from the inclination $i$ of the orbit, the angle of the line of nodes $\Omega$ and the argument of the pericentre $\omega$, via
\begin{align}
    & \mathcal{A}=\cos\Omega\cos\omega-\sin\Omega \sin\omega \cos i\,,\\
    & \mathcal{B}=\sin\Omega \cos\omega+\cos\Omega \sin\omega \cos i\,,\\
	& \mathcal{C}=-\sin\omega \sin i\,,\\
    & \mathcal{F}=-\cos \Omega \sin\omega-\sin\Omega \cos\omega \cos i\,,\\
    & \mathcal{G}=-\sin\Omega\sin\omega+\cos\Omega \cos\omega \cos i\,,\\
    & \mathcal{H}=-\cos\omega \sin i\,.
\end{align}
A similar expression leads from the orbital velocities $(v_{\rm x}, v_{\rm y})$ to the sky-projected ones $(V_X, V_Z)$ and the radial velocity $V_Z$:
\begin{align}
 	V_X&= \mathcal{B}v_{\rm x}+\mathcal{G}v_{\rm y}\,,\\
 	V_Y &= \mathcal{A}v_{\rm x}+\mathcal{F}v_{\rm y}\,,\\
	V_Z &= -(\mathcal{C}v_{\rm x}+\mathcal{H}v_{\rm y})\,.
\end{align}
Additionally, in order to properly reconstruct synthetic orbits for S2, we have to take into account on the projected quantities the classical Rømer delay and the redshift on the radial velocity coming from special relativity and gravitational time dilation (we refer to \citep{Grould2017, DellaMonica_2021, deMartino_2021} for a detailed analysis on how these effects can be quantified on our predicted orbits). Finally, we correct our synthetic orbits for systematic effects arising from the construction of the reference frame. These result in 5 additional parameters in the generation of a synthetic orbit of S2, namely: ($x_{\bullet}$,$y_{\bullet}$) related to a zero-point offset of the central mass w.r.t. the origin of the astrometric reference frame; ($v_{x,\bullet}$,$v_{y,\bullet}$) describing a drift over time of the central mass on the astrometric reference frame; $v_{z,\bullet}$ related to an offset in the estimation of the radial velocity of the S2 star when correcting to the Local Standard of Rest.

%%%%%%%%%%%%%%%%%%%%%%%%%%%%%%%%%%%%%%%%%%%%%%%%%%

% Don't change these lines
\bsp	% typesetting comment
\label{lastpage}
\end{document}

\begin{figure*}
    \centering
    \includegraphics[width = \textwidth]{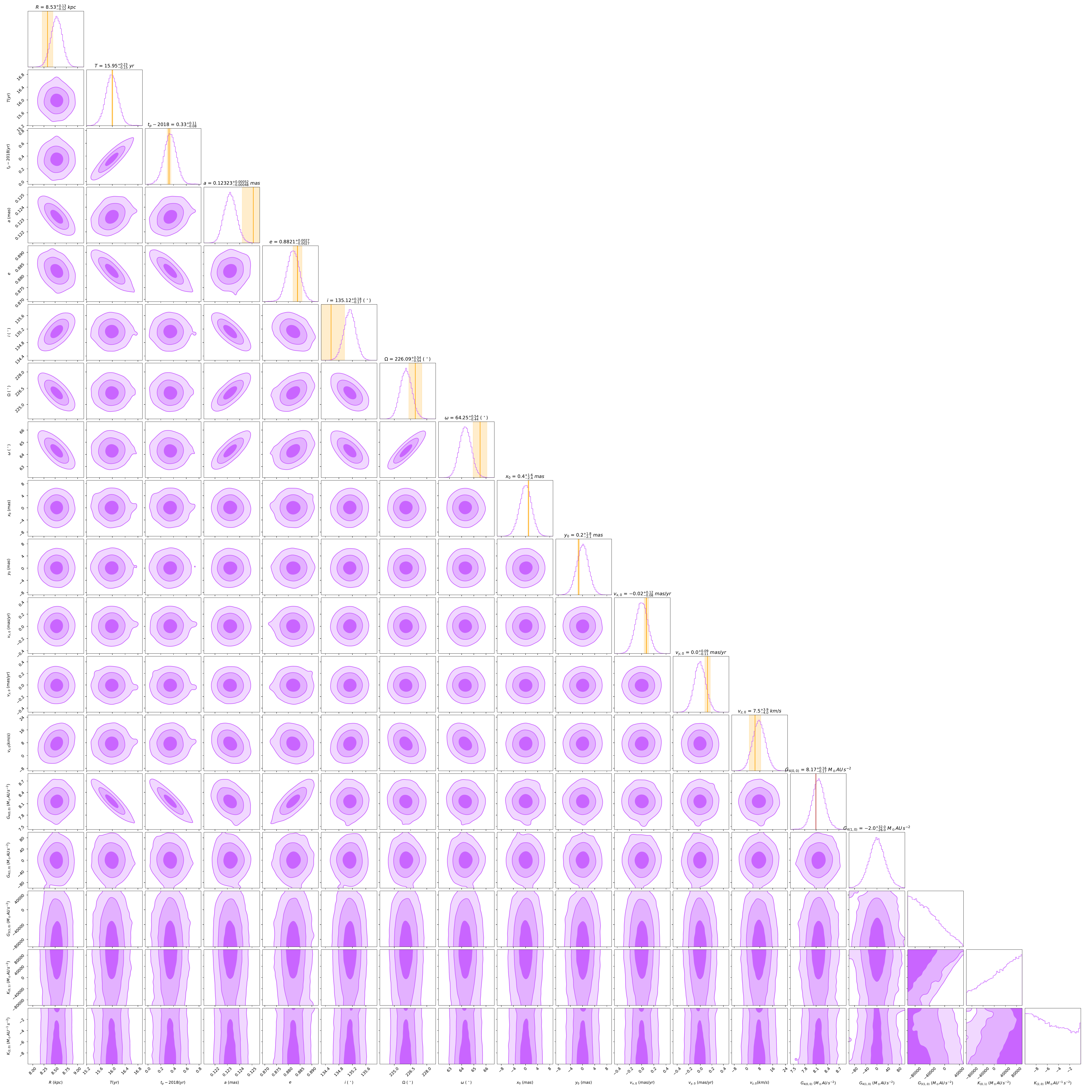}
    \caption{The figure illustrate the $1\sigma$,$2\sigma$, and $3\sigma$ contours of the full posterior distribution of the Newtonian limit of the  Horndeski gravity. The parameters $G_{4(0,0)}$ and $G_{4(1,0)}$ are bound and the mean value of their distributions, plus the 68\% confidence interval are reported on the respective histograms. The other parameter are not constrained and  for them we can only infer exclusion regions in the parameter space. Red lines and the associated shaded regions represent the values of the parameters obtained by Gillessen et al (2017).}
    \label{fig:mcmc_posterior_full}
\end{figure*}